# The Nature of G52.381-0.849 and G56.240-0.345: YSOs associated with Extended MIR Emission?

J.P. Phillips, J.A. Pérez-Grana,
G. Ramos-Larios, S. Velasco-Gas

Instituto de Astronomía y Meteorología, Av. Vallarta No. 2602, Col. Arcos Vallarta, C.P. 44130 Guadalajara, Jalisco, México   e-mail : jpp@astro.iam.udg.mx

**Abstract**

We report the results of visual spectroscopy, mid-infrared (MIR) mapping and photometry, and near infrared (NIR) photometry of two candidate symbiotic stars (IPHAS J193108.67+164950.5 and IPHAS J193709.65+202655.7) associated with extended MIR emission. Our analysis of the continua of these sources shows that they are likely to represent Class I-II young stellar objects (YSOs) in which most of the IR emission arises from circumstellar disks, and for which the physical characteristics (stellar temperatures, radii, masses and luminosities) are similar. The extended emission is characterised by a substantial increase in fluxes and dimensions to longer MIR wavelengths. This is likely to arise as a result of emission by polycyclic aromatic hydrocarbons (PAHs) within extended photodissociation regimes (PDRs), centred upon more compact ionized regions responsible for much of the shorter wave emission. Such dual emission structures are characteristic of those observed in many compact HII regions. Finally, we note that the clouds have asymmetrical structures and wind-swept morphologies, conceivably indicative of shock interaction with external winds. Where this is the case, then it is possible that the YSOs are located in regions of triggered star-formation.

**Key Words:** (ISM:) HII regions --- stars: formation --- (stars:) circumstellar matter --- (stars:) binaries: symbiotic --- stars: pre-main-sequence --- ISM: lines and bands

# 1. Introduction

The INT Photometric H$\alpha$ survey of the Northern Galactic Plane (IPHAS; Drew et al. 2005) has resulted in the discovery of many new candidate planetary nebulae (Viironen et al. 2009a, b), AGB stars (Wright et al. 2009), and very low mass accreting sources (Valdivieso et al. 2009) among other objects, and promises to reveal much further information concerning the structure and contents of our Galaxy. Corradi et al. (2008) have also noted that a judicious combination of 2MASS near infrared (NIR) and IPHAS results can be used to identify candidate symbiotic stars; an analysis which has the potential of greatly of increasing the corpus of such sources. An investigation of these sources using the first Galactic Legacy Infrared Midplane Survey Extraordinaire (GLIMPSEI), an MIR mapping program undertaken with the *Spitzer Space Teles*cope (*SST*; Werner et al. 2004), has permitted us to determine that three of the sources are centred within compact regions of emission. Although none of the nebulae have structures similar to those observed in other symbiotic sources (see e.g. Corradi et al. 1999), they are nevertheless clearly associated with the stars, and merit further observations. It is conceivable that they represent nebular outflows about D-type symbiotics, for which only a relatively few cases (~24) are at present known.

We report here the investigation of two of the sources, G52.381-0.849 and G56.240-0.345, centred close to the symbiotic candidates IPHAS J193108.67+164950.5 and IPHAS J193709.65+202655.7 (referred to hereafter as IPHAS-Sy745 and IPHAS-Sy750 – sources 745 and 750 in the table of candidate symbiotics listed by Corradi et al 2008. The case of the third source IPHAS J195001.48+262738.1 $\equiv$ IPHAS-Sy774 has been discussed separately by Ramos-Larios, Phillips, & Perez-Grana 2009). Neither of these regions have been observed at other wavelengths, and they appear to represent previously unknown examples of emission or reflection envelopes.

We shall discuss the structures of these envelopes in the MIR; present visual spectroscopy of the candidate symbiotic stars; and undertake modelling of the visual-infrared continua of the stars in an attempt to understand the nature of the sources.

We shall conclude, as a result, that both of these objects are likely to represent young stellar objects (YSOs) cocooned within the remnants of the clouds out of which they were born. Although the nature of the enveloping shells is far from well established, it seems likely that they represent compact HII regions enveloped by photo-disssociative regimes (PDRs), wherein much of the MIR emission derives from polycyclic aromatic hydrocarbons (PAHs). It is also possible that their asymmetrical structures derive from interaction with external winds, and that the YSOs arise from a process of triggered star-formation.

## 2. Observations

Low dispersion optical spectra of IPHAS-Sy745 and IPHAS-Sy750 were obtained on the 15$^{th}$ of May 2009, using a Boller & Chivens spectrometer mounted at the prime focus of the 2.1 m telescope, Observatorio Astronomico Nacional (OAN-SPM), San Pedro Martir, Baja California, Mexico. A CCD SITe1 1024x1024 CCD was used as detector, in tandem with a grating having 300 l/mm set and a blaze of 5000 Å. The slits had lengths of 5 arcmins, and widths of 180 $\mu$m ($\equiv$ 2.35 arcsec), whilst the plate and spectral scales were 1.05 arcsec pix$^{-1}$ and 2.36 Å pix$^{-1}$ respectively. The spectral resolution was 10.66 Å, the wavelength uncertainty of order ~4 Å, and the spectral range was 3500 Å$\rightarrow$7200 Å - although the limits of this range have a lower S/N, and have been excluded from the present analysis. Three exposures of 15 min each were obtained for IPHAS-Sy745, to give a combined total of 45 min, whilst 2x15 min exposures were acquired for IPHAS-Sy750. Mean air masses were low and of order 1.2$\rightarrow$1.3 for IPHAS-Sy745, and 1.01 for IPHAS-Sy750, whilst the results were calibrated using the stars Hz 44, Feige 34 and BD+362242.

Mid-infrared (MIR) mapping of the associated emission regions G52.381-0.849 and G56.240-0.345 was obtained using data deriving from the Galactic Legacy Infrared Midplane Survey Extraordinaire ("GLIMPSE"), wherein 220 square degrees the galactic plane was surveyed at a pixel size of 0.6 arcsec, and to a pointing accuracy of ~ 0.3 arcsec. The observations were undertaken using the Infrared Array Camera (IRAC; Fazio et al. 2004), and employed filters having isophotal wavelengths (and bandwidths $\Delta\lambda$) of 3.550 $\mu$m ($\Delta\lambda$ = 0.75 $\mu$m), 4.493 $\mu$m ($\Delta\lambda$ = 1.9015 $\mu$m), 5.731 $\mu$m ($\Delta\lambda$ = 1.425 $\mu$m) and 7.872 μm (Δλ = 2.905 μm). The normal spatial resolution for this instrument varies between ~1.7 and ~2 arcsec (Fazio et al. 2004), and is reasonably similar in all of the bands, although there is a stronger diffraction halo at 8 μm than in the other IRAC bands. This leads to differences between the point source functions (PSFs) at ~0.1 peak flux.

Contour maps have been produced in the differing IRAC bands, where the emission $E_n$ for contour level $n$ is given through $E_n = A10^{(n-1)B}$ MJy sr$^{-1}$, and contour parameters *[A,B]* are cited in the captions to Figs. 1 & 2. Similarly, profiles through the sources are given in Fig. 3, a process which required correction for the effects of background emission in the differing photometric bands. These components are particularly strong at 5.8 and 8.0 μm, where they also result in slight gradients of order of 5 10$^{-4}$ MJy/sr/pix (although these gradients also depend upon the direction of the slice). We have removed these trends by subtracting lineal ramps from the results – a procedure which is more than adequate given the limited size of the sources.

Some care must be taken in interpreting the profiles, however. The problems with large aperture photometry are described in the IRAC data handbook, and relate in part to scattering in an epoxy layer between the detector and multiplexer (Cohen et al. 2007). This leads to the need for flux corrections as described in Table 5.7 of the handbook; corrections which are of maximum order 0.944 at 3.6 μm, 0.937 at 4.5 μm, 0.772 at 5.8 μm and 0.737 at 8.0 μm. However, the precise value of this correction also depends on the underlying surface brightness distribution of the source, and for objects with size ~several arcminutes it is counselled to use corrections which are somewhat smaller. The handbook concludes that "this remains one of the largest outstanding calibration problems of IRAC".

We have, in the face of these problems, chosen to leave the profiles unchanged. The maximum correction factors for the 8.0μm/4.5μm and 5.8μm/4.5μm ratios are likely to be > 0.8, but less than unity, and ignoring this correction has little effect upon our interpretation of the results.

Finally, the 2MASS all-sky survey was undertaken between 1997 and 2001 using 1.3 m telescopes based at Mt Hopkins, Arizona, and at the CTIO in Chile, and resulted in photometry of > 5 $10^8$ sources in the J (1.25 μm), H (1.65 μm) and $K_S$ (2.17 μm) photometric bands (Skrutskie et al. 2006). We have used results from the point source catalogue to obtain fluxes for both the candidate symbiotics, and a likely Class I YSO located close to the centre of G56.240-0.345. These are used to constrain the Spectral Energy Distributions (SEDs) of the sources, and to limit the range of YSO models evaluated by Robitaille et al. (2006).

## 3. The structures of G52.381-0.849 and G56.240-0.345

Contour mapping of of G52.381-0.849 and G56.240-0.345 is illustrated in Figs. 1 & 2, wherein the positions of the candidate symbiotic stars of Corradi et al. (2008) are indicated using solid white circles, and that of a likely bright central YSO is denoted using a dashed white circle. It is, apparent from these results, that the envelopes are extremely compact at 3.6 and 4.5 μm (size ≈ 30 arcsec), but greatly increase in size to longer IRAC wavelengths. This trend is very similar to what is observed in compact HII regions (e.g. Phillips & Ramos-Larios 2008a) and planetary nebulae (e.g. Phillips & Ramos-Larios 2008b), and is usually interpreted in terms of a dual emission structure. Whilst short wave emission tends to be dominated by permitted and forbidden line fluxes, molecular transitions and bremsstrahlung continuum, much of which arises in ionised plasmas in the inner portions of the sources, the more extended and much stronger emission noted at longer IRAC wavelengths is likely to derive from warm dust continua, and PAH emission bands at 3.3, 6.2, 7.7 and 8.6 μm (see e.g. Tielens 2005, and the spectra published by Peeters et al. 2002).

These trends in emission are also apparent in the MIR profiles illustrated in Fig. 3, wherein we illustrate slices across the centres of the sources and the two candidate symbiotic stars. It is apparent from these that the candidate symbiotics appear strong at all of the MIR wavelengths (the stars are located at a relative position (RP) of 0 arcsec for G52.381-0.849, and at RP = -19 arcsec for G56.240-0.345), as is the case for the central source in G56.240-0.345 (located at RP = 0 arcsec). An analysis of MIR point source photometry, allied to the theoretical colour-colour analysis of Allen et al. (2004), suggests that the latter object represents one of only two Class I YSOs within the field of G56.240-0.345, and has colours [3.6]-[4.5] = 1.01 and [5.8]-[8.0] = 1.44. The second source with Class I MIR colours is the candidate symbiotic itself, for which we find [3.6]-[4.5] = 0.646 and [5.8]-[8.0] = 1.498. It is always possible that further such sources remain to be detected, however. Many YSOs may be located close to or below the threshold of detection, and are either excluded from the point source catalogue, or have incomplete photometry.

Finally, we note that there appear to be no Class I sources within the field of G52.381-0.849, although we shall argue below that the symbiotic candidate IPHAS-Sy745 is likely to represent a more evolved species of YSO.

Apart from this, it is clear that the short wave 3.6 and 4.5 $\mu$m emission trends are very closely similar. In the case of G56.240-0.345, for instance, it is apparent that emission is confined between RPs of -30 and +15 arcsec. A similar trend applies for G52.381-0.849, where it is apparent that short wave emission is confined to a range of ~30 arcsec. Although it is possible that 3.3 $\mu$m PAH band features are contributing to the 3.6 $\mu$m results, the similarity between the 3.6 and 4.5 $\mu$m trends is suggestive of ionised and stellar components of emission.

The 5.8 and 8.0 $\mu$m components, by contrast, are very much more extended, and significantly stronger, with the 8.0$\mu$m/4.5$\mu$m and 5.8$\mu$m/4.5$\mu$m ratios taking values of order ≈20 and ≈7.5 respectively. It is also apparent that emission in these channels is very much more extended, and likely arises from neutral PDRs such as have been described by Tielens (2005), Hollenbach & Tielens (1997) and Abel et al. (2005) (see also Phillips & Ramos-Larios 2008a, b). Conditions in these regions are much more conducive to the survival of small PAH emitting grains, given that temperatures may be of order <130 K (e.g. Abel et al. 2005), and UV radiation field energy densities are reduced. By contrast, the much more hostile environment of the central HII regions is likely to result in the rapid sublimation and sputtering of the grains (e.g. Allain, Leach & Sedlmayr 1996a, 1996b).

Finally, it is apparent that the maps in Figs. 1 & 2, and the profiles in Fig. 3 are strongly asymmetric. In the case of the profiles for G52.381-0.849, for instance, it is apparent that MIR emission rises steeply between RPs of ~15 and 5 arcsec; peaks at the position of the symbiotic candidate; and falls-off more slowly towards negative RPs. A similar, although somewhat less extreme tendency is also found in G56.240-0.345, and is apparent in the asymmetric distributions of emission noted in Figs. 1 & 2. The overall impression, from the observed morphologies and profiles, is that the regions are being compressed on their ~southerly (in the case of G52.381-0.849) and ~easterly (G56.240-0.345) sides, and that this is leading to cometary tails of wind-swept material, and the diffuse appearances in Figs. 1 & 2. It is unclear, at present, what the origins the external winds or radiation pressure fields might be – there are for instance no recorded supernova remnants within the relevant areas of sky (viz. the recently updated catalog of SNRs published by Green (2009), available at the Vizier web page http://cdsarc.u-strasbg.fr/viz-bin/Cat?VII/253). It is entirely possible, however, that impacting winds are shock-interacting with the globules, and stimulating the formation of the observed stars (and likely YSOs – see our later discussions in Sect. 4) – a case of triggered star formation similar to that observed in other star formation centres (see e.g. the discussion of Phillips, Ramos-Larios & Perez-Grana 2009; Boss 1995; Foster & Boss 1996, 1997; Nakamura et al. 2006; Melioli et al. 2006; Boss et al. 2008). It would be of interest to investigate the clouds over a fuller range of wavelengths, and in the context of the ISM environment within which they are located.

**4. The Stars in G52.381-0.849 and G56.240-0.345**

We have obtained spectra for the symbiotic candidates in G52.381-0.849 (IPHAS-Sy745) and G56.240-0.345 (IPHAS-Sy750) using the B&Ch spectrograph described in Sect. 2. The results are illustrated in Fig. 4. It is clear, from these, that there is no evidence for H$\alpha$ emission within the sources – a tendency which is rather surprising given the r-H$\alpha$ colours quoted by Corradi et al. (2008). It is also apparent that whilst TiO absorption appears to be present in IPHAS-Sy745, there is no such evidence for the case of IPHAS-Sy750.

Finally, we see no evidence for forbidden lines such as [OIII] $\lambda$5007 Å, or higher excitation transitions such as HeII $\lambda$4686 Å. The absence of these transitions suggests that neither of these sources is a bona-fide symbiotic star. On the other hand, and as is apparent in Figs. 1→3, both of the stars appear strong at MIR wavelengths, and may represent examples of YSOs.

Given this possibility, we have combined V band photometry determined from our present spectroscopy; J, H & $K_S$ band photometry from 2MASS; and MIR photometry from the *SST* in an attempt to define the visual-IR continua of the sources. We shall use these trends to determine whether the stars can be plausibly modelled in terms of YSO continua and, if so, what the properties of the sources might be.

**4.1 Modelling of IPHAS-Sy745**

Photometry for IPHAS-Sy745 has been analysed using the ~2 $10^5$ YSO models evaluated by Robitaille et al. (2006), together with the SED fitting tool developed by Robitaille et al. (2007), in which continua have been determined for differing dust and gas geometries, varying dust properties, and using 2D radiative transfer modelling for a large region of parameter space. The fitting of the models also takes account of the apertures employed for the photometry. Given that our present source does not appear to be resolved in any of the wavebands, we have set this parameter to be equal to the FWHM of the PSFs in each of the respective channels – ranging from ~2 arcsec for the IRAC and visual wavebands, to ~5 arcsec for the 2MASS results.

Finally, since neither the distance nor interstellar extinction of this source are very well known, we have used reasonably broad ranges of values for both of these parameters: $0.01 < D/kpc < 10$ for distance D, and $0 < A_V/mag < 50$ for extinction $A_V$. Identical ranges will be employed for modelling of the stars in G56.240-0.346 (see Sect. 4.2).

The results of this analysis are illustrated in Fig. 5, where we show the photometric results in the upper left-hand panel, together with associated model fits. The black curve corresponds to the best fit model, for which $\chi^2 \equiv \chi_{BEST}^2 = 0.89$, whilst the grey curves correspond to models in which $\chi^2 - \chi_{BEST}^2 < 3$. The dashed curve, finally, indicates the variation of photospheric emission from the star, for which we have included the effects of interstellar extinction, but excluded those of circumstellar extinction.

The individual components making up the best-fit model continuum are illustrated in the upper right-hand panel, wherein the total flux is indicated as black; the stellar flux is blue; the stellar photospheric flux is indicated by the dashed line (this is the flux prior to reddening by circumstellar dust); the disk flux is green; the scattered flux is yellow; the envelope flux is red; and the thermal flux is orange. Unless otherwise stated, the results include the effects of circumstellar extinction, but not of IS extinction. They also assume a representative distance of 1 kpc, and an aperture of close to 5 arcsec.

It is plain from this that the combined visual, NIR and MIR continuum can be very well modelled in terms of a Class II YSO, for which the predominant emission derives from the stellar disk and central star. The contribution of the larger envelope is poorly defined, since we have no longer-wave photometry with which to adequately constrain this component. However, there is no particular reason to suppose that it contributes appreciably to observed or longer wave IR fluxes.

Histograms of various of the physical parameters associated with these models are illustrated in Fig. 6, where the grey bars indicate the relative numbers of parameters searched in the present analysis. The cross-hatched areas show results for the best-fit ($\chi^2 - \chi_{BEST}^2 < 3$) model solutions. It is plain, from this, that the characteristics of the YSO are tolerably well defined. The large majority of models indicate a luminosity range ~50→$10^3$ $L_\odot$, masses ~30→60 $M_\odot$, stellar radii ~2→4 $R_\odot$, and temperatures $10^4$→2 $10^4$ K. The age is likely to be on the order of several millions of years (the large majority of models are concentrated in this regime), whilst interstellar extinction $A_V$ is of order 4-6 mag, and extinction due to circumstellar dust (not shown here) is likely to be modest (< 1 mag). Despite the inevitable uncertainties in any such an analysis, it is therefore clear that we are dealing with a reasonably high mass and luminosity star.

### 4.2 Modelling of the Stars in G56.240-0.345

Corresponding modelling for G56.240-0.345 is illustrated in the lower panels of Fig. 5, where we show results for the bright central star (and probable Class I YSO - Star A, central panels) and IPHAS-Sy 750 (Star B, lower panels). Here again, and as for the case of the central star in G52.381-0.849, it is clear that fluxes continue to increase towards the MIR, and are capable of being fitted using model YSO continua. Whilst the continua of both of the sources are dominated by disk and stellar components of emission, it is possible that they are associated with appreciable longer wave emission deriving from the larger envelope component; a result which is consistent with the Class I colour indices cited in Sect. 3.

The range of physical parameters associated with the central YSO is illustrated in Fig. 7, where we have plotted all of these values against the interstellar extinction $A_V$. Given that neither of these regions has been detected in the NIR or visible, it is probably fair to assume that extinctions are appreciable and of order $A_V > 5$ mag. This would imply, for most of the models, that masses $M_*$ are of order $\sim 2 \rightarrow 10$ $M_\odot$, luminosities $L_*$ are in the range $\sim 30 \rightarrow 4000$ $L_\odot$, ages are of order several millions of years, stellar temperatures $T_{EFF}$ are $\sim 10^4 \rightarrow 2 \cdot 10^4$ K, photospheric radii are $\sim 2 \rightarrow 6$ $R_\odot$, and that circumstellar extinctions are small.

IPHAS-Sy750, on the other hand, is associated with a much smaller range of model fits, although if one loosens the fit criteria such that $\chi^2 - \chi^2_{BEST} < 5$, then it is clear that stellar masses would be of order 4-10 $M_\odot$, effective temperatures $\sim 2 \cdot 10^4$ K, photospheric radii of order $\sim 3$ $R_\odot$, luminosities $\sim 3 \cdot 10^3$ $L_\odot$, and the age $\sim 3 \rightarrow 7 \cdot 10^5$ yrs. All of the latter values are calculated by assuming larger values of $A_V$ – although there are no solutions having $A_V > 3$ mag.

So it is clear that all of these stars are likely to represent YSOs associated with low-mass HII regions - none of them is associated with the symbiotic phenomenon; a conclusion which also applies for the source IPHAS J195001.48+262738.1 investigated by Ramos-Larios, Phillips & Perez-Grana (2009). Whilst this subset of sources is too small to come to statistically significant conclusions, it suggests that a large fraction of the sample of Corradi et al. (2008) may have been mis-identified. Such a conclusion is also consistent with the low detection rates for D-type symbiotics noted by Corradi et al. (2009), and the high levels of contamination previsioned by Corradi et al. (2008).

It is also apparent that the modelling of Robitaille et al. (2006) permits us to synthesise the continua exceptionally well. Whilst there are some disparities in the properties of the differing stars, one is struck by the level of overlap in their physical characteristics. All of them have masses which are appreciable (of order ~3→10 $M_\odot$), have high luminosities, large stellar radii, and temperatures of order ≈$10^4$ K. It is also clear that the MIR and NIR continua are dominated by the stellar disks, whilst visual fluxes correspond to the highly extincted photospheric continua.

## 5. Conclusions

We have investigated two candidate symbiotic stars of Corradi et al. (2008) associated with extended MIR emission. Visual spectroscopy of the stars shows that whilst one of them (the source IPHAS-Sy745) is associated with TiO absorption bands, neither of them have the higher excitation or forbidden lines which are normally characteristic such stars. On the other hand, both of the stars have steeply rising visual-IR continua, and these have proven susceptible to YSO modelling using the 2D radiative transfer codes of Robitaille et al. (2006). The results suggest that the stars may represent Class I→II YSOs in which most of the MIR/NIR emission derives from circumstellar disks.

We also note that the MIR colour indices of the central star in G56.240-0.345 are indicative of a Class I YSO, and although we have no visual spectroscopy for this source, we are again able to synthesise the continuum using the Robitaille et al. models.

The physical properties of these sources, as determined from this analysis, imply that the characteristics of the stars are somewhat different. They agree however in suggesting that masses and luminosities are large (respectively ≈2→60 $M_\odot$ and 30→4000 $L_\odot$), temperatures are of order ~1→2 $10^4$ K, and that photospheric radii are in the range 2→6 $R_\odot$.

The extended MIR emission may derive from remnants of the gas out of which the stars were formed, and shows evidence for higher levels of emission, and more extended structures in the longer wave 5.8 and 8.0 $\mu$m IRAC bands. These trends are likely to arise due to PAH emission within the photo-dissociative regimes. By contrast, the regimes of shorter wave emission are much more compact, and may be associated with ionised gas within the inner sectors of the shells. Such dual emission structures, where emission derives from ionised cores and more extended neutral envelopes, are characteristic of the structures observed in many compact HII regions. It is also noted that the shells have a wind-swept appearance, and appear to be compacted on one side of their shells; an indication perhaps of interaction with an external wind, and triggered star-formation within the sources.

**Acknowledgements**

This research is based, in part, on observations made with the Spitzer Space Telescope, which is operated by the Jet Propulsion Laboratory, California Institute of Technology under a contract with NASA. Support for this work was provided by an award issued by JPL/Caltech. In addition to this, the work makes use of data products from the Two Micron All Sky Survey, which is a joint project of the University of Massachusetts and the Infrared Processing and Analysis Center/California Institute of Technology, funded by the National Aeronautics and Space Administration and the National Science Foundation. The 2MASS data was acquired using the NASA/ IPAC Infrared Science Archive, which is operated by the Jet Propulsion Laboratory, California Institute of Technology, under contract with the National Aeronautics and Space Administration. GRL acknowledges support from CONACyT (Mexico) grant 93172.

# Figure Captions

**Figure 1**

MIR contour mapping for source G52.381-0.849, where the contour parameters [A,B] are given by [2.3, 0.1851] at 3.6 $\mu$m, [2.0, 0.1906] at 4.5 $\mu$m, [10.0, 0.1470] at 5.8 $\mu$m, and [25.0, 0.1141] at 8.0 $\mu$m. Note how the dimensions of the source increase markedly towards larger MIR wavelengths – a trend which is likely to arise due to increasing levels of PAH emission within the enveloping PDRs. We have indicated the candidate symbiotic source of Corradi et al. (2008) by means of a white circle within the 3.6 $\mu$m map.

**Figure 2**

As for Fig. 2, but for the source G56.240-0.345. In this case, the contour parameters [A,B] are given by [2.2, 0.1667] at 3.6 $\mu$m, [1.8, 0.1746] at 4.5 $\mu$m, [8.0, 0.1591] at 5.8 $\mu$m, and [21.0, 0.1252] at 8.0 $\mu$m. We have indicated the candidate symbiotic source of Corradi et al. (2008) using a white solid circle, and a likely Class I YSO using the white dashed circle (see the 3.6 $\mu$m map).

**Figure 3**

MIR profiles through the sources G52.381-0.849 (upper panel) and G56.240-0.345 (lower panel), where the directions and widths of the slices are indicated in the inserted images. The candidate symbiotic star for G52.381-0.849 (IPHAS-Sy745) is located at RP = 0 arcsec. By contrast, the corresponding source in G56.240-0.345 (IPHAS-Sy750) is positioned at RP = -19 arcsec, and the peak at RP = 0 arcsec corresponds to a probable Class I YSO. Note, for both of these cases, how source fluxes and sizes increase with increasing IRAC wavelength.

**Figure 4**

Spectra for the two candidate symbiotic stars IPHAS-Sy745 and IPHAS-Sy750. Note the presence of TiO absorption bands in the uppermost spectrum.

**Figure 5**

YSO modelling of the visual-IR continuum for the central star in G52.381-0.849 (IPHAS-Sy745), and two stars associated with G56.240-0.345; where in the latter case, the central YSO is labelled as star A, and IPHAS-Sy750 corresponds to star B. The left-hand panels show the fitting of our results using YSO continua deriving from Robitaille et al. (2006), whilst the right-hand panels show the differing flux components for the best fit models, determined for a representative distance of 1 kpc, and an aperture size of ~5 arcsec. It is clear that most of the NIR + MIR emission derives from circumstellar disks (indicated using the green curves). The curves in the right-hand panels do not take account of interstellar extinction, and this explains the apparent differences between these trends and those illustrated in the left hand panels (see also the discussion in Sect. 4.1).

**Figure 6**

Results of YSO continuum modelling for the central star in G52.381-0.849, where we show histograms for six representative stellar physical parameters. Grey areas indicate the relative numbers of models investigated in the analysis, whilst cross-hatched regions correspond to the distributions of best fit models. Both series of results are normalised to unity. It is clear that the stellar luminosity and mass are likely to be appreciable.

**Figure 7**

Parametric solutions for the central star in G56.240-0.345, based upon YSO modelling of the visual-IR continuum. All of the parameters are represented against the interstellar extinction $A_V$.

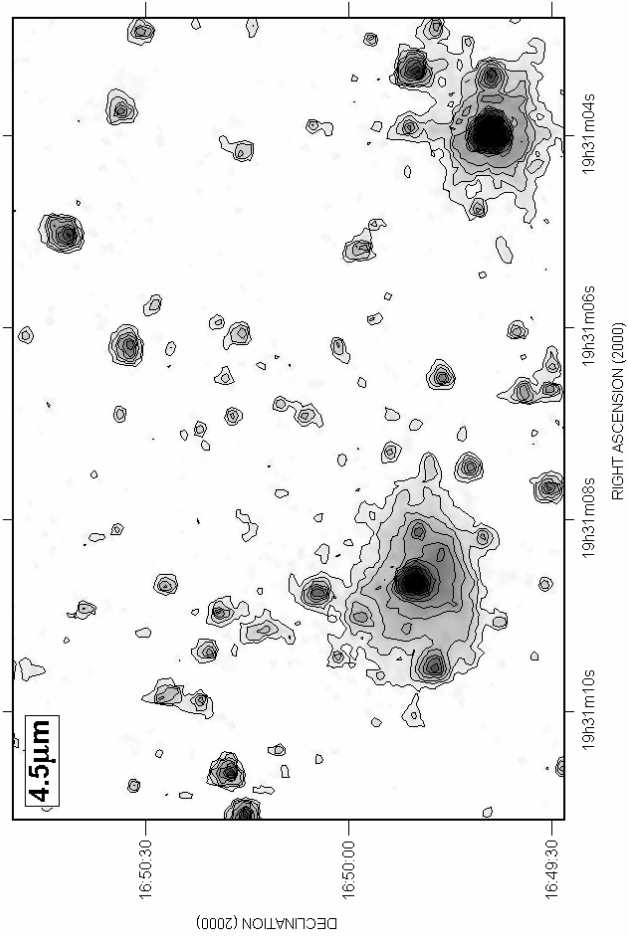
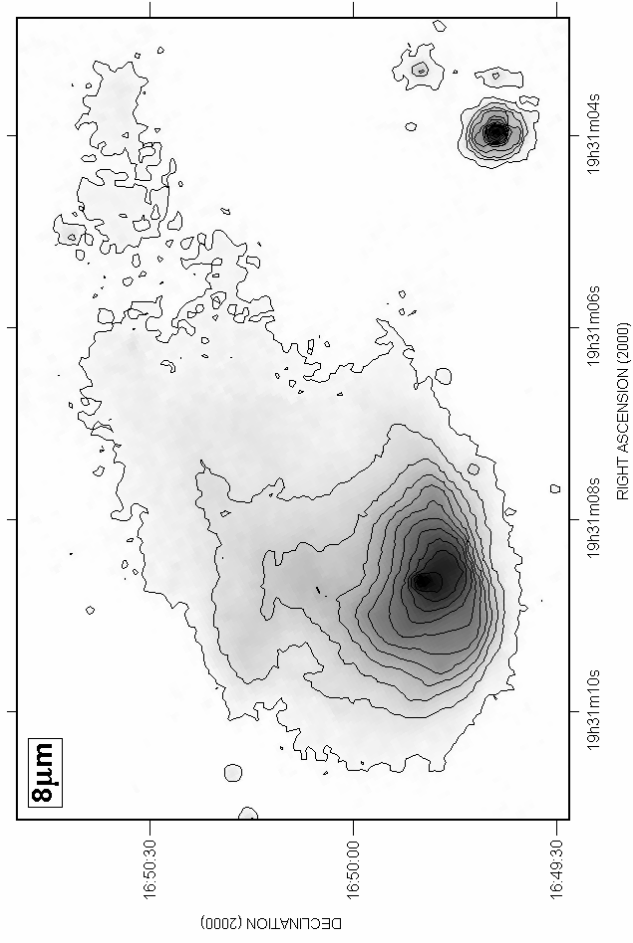
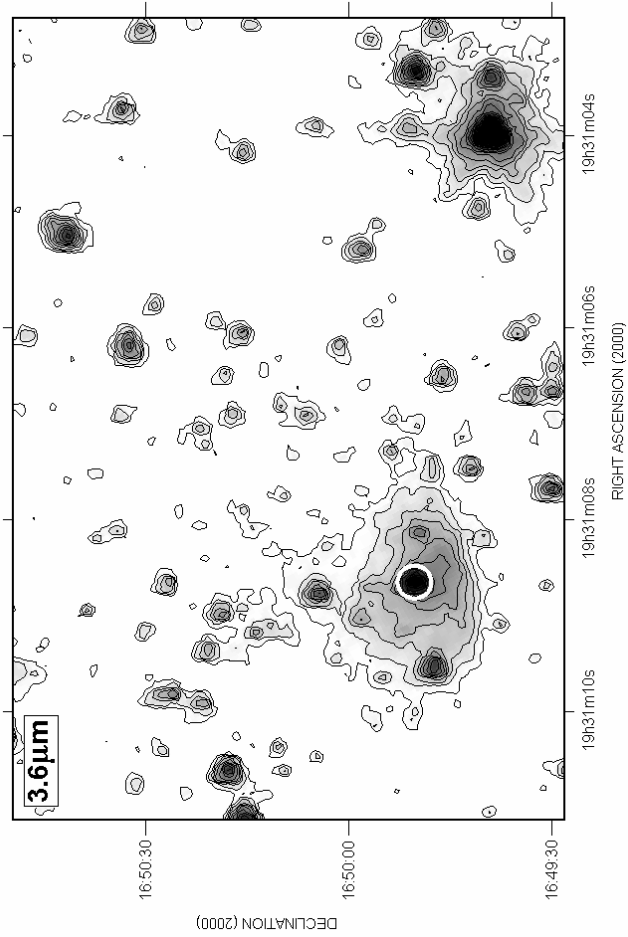
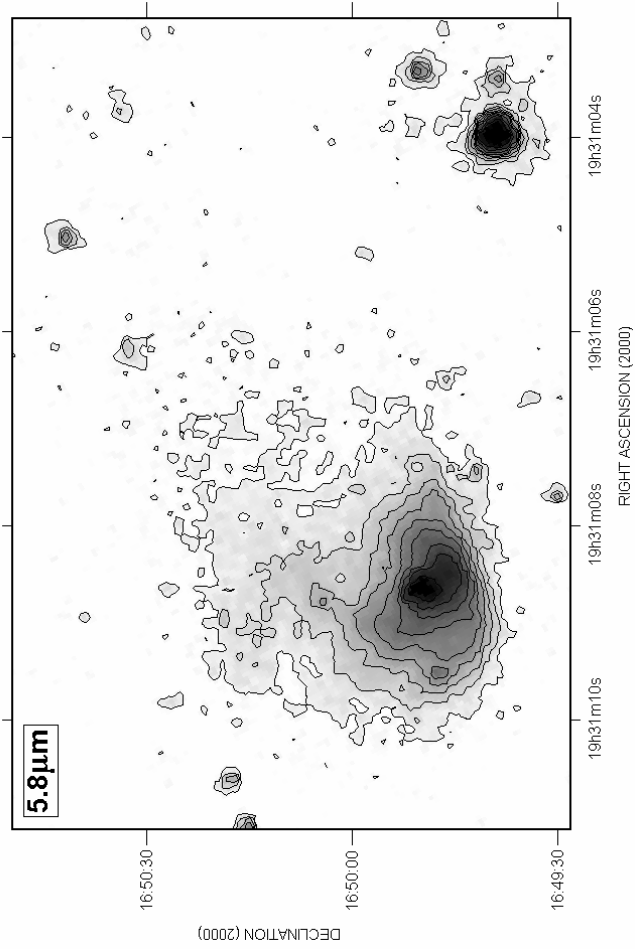

FIGURE 1

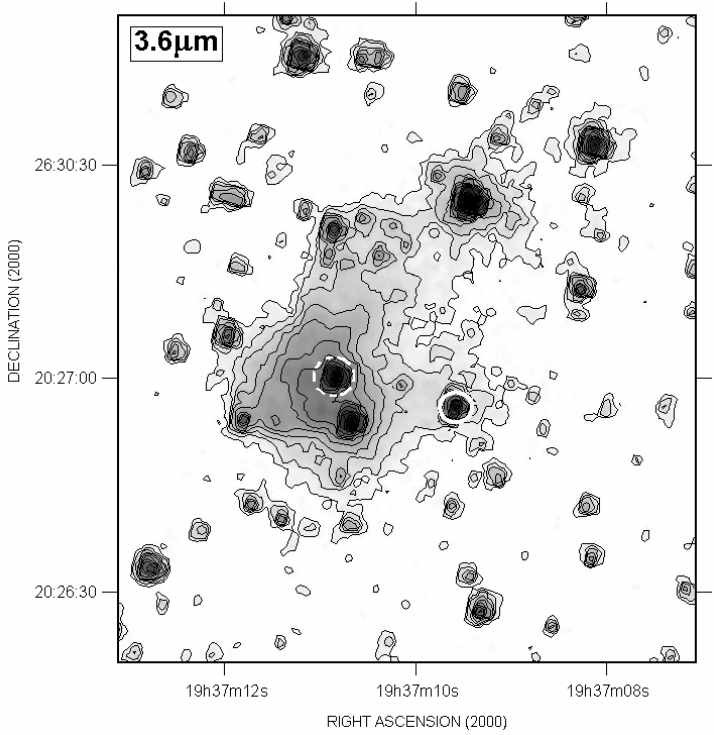 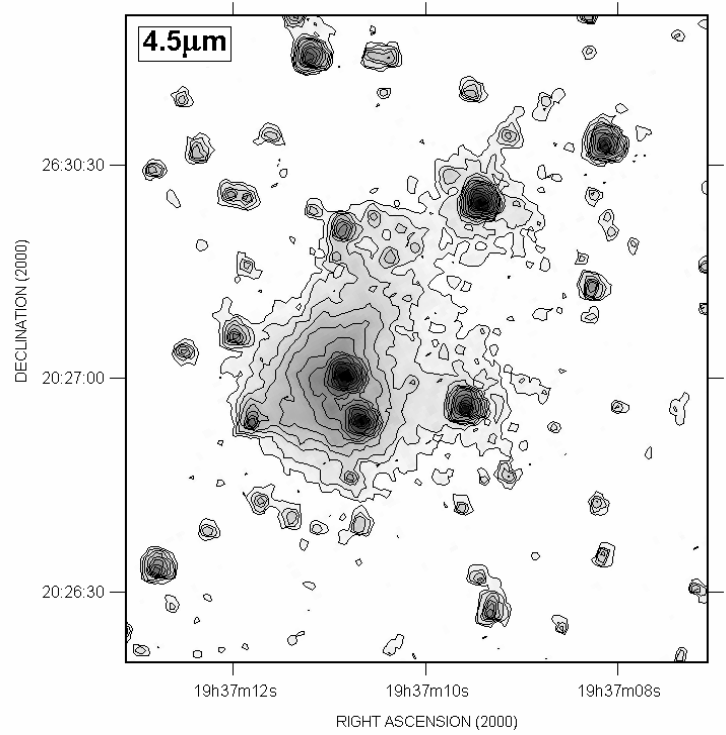
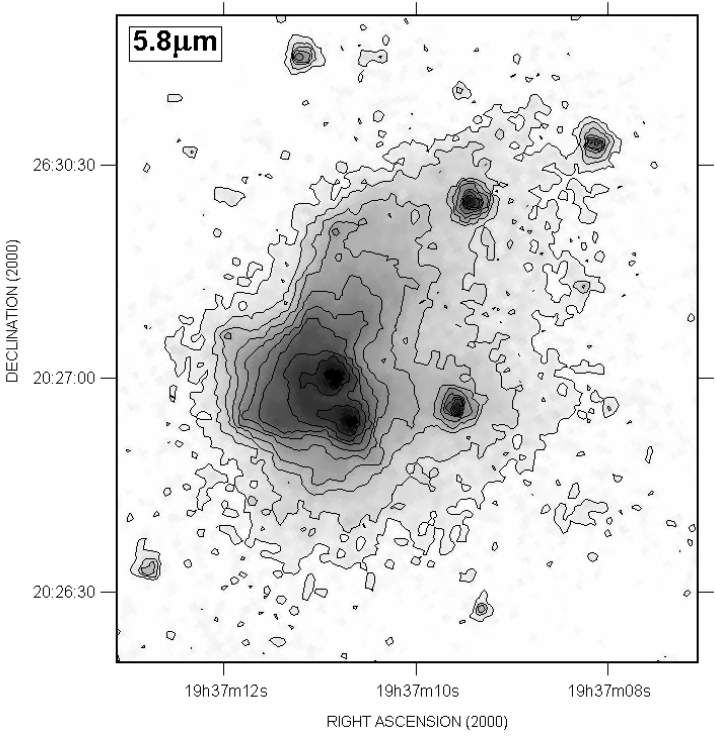 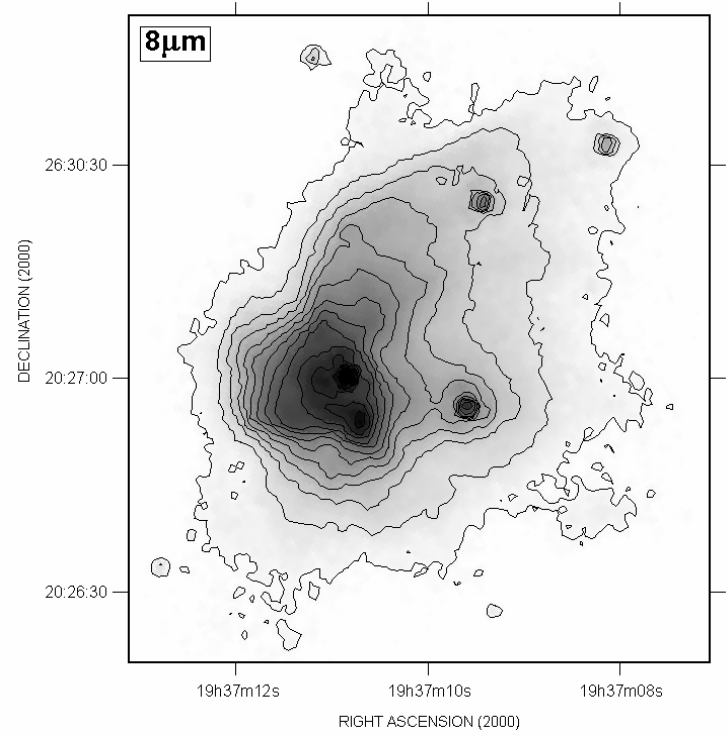

FIGURE 2

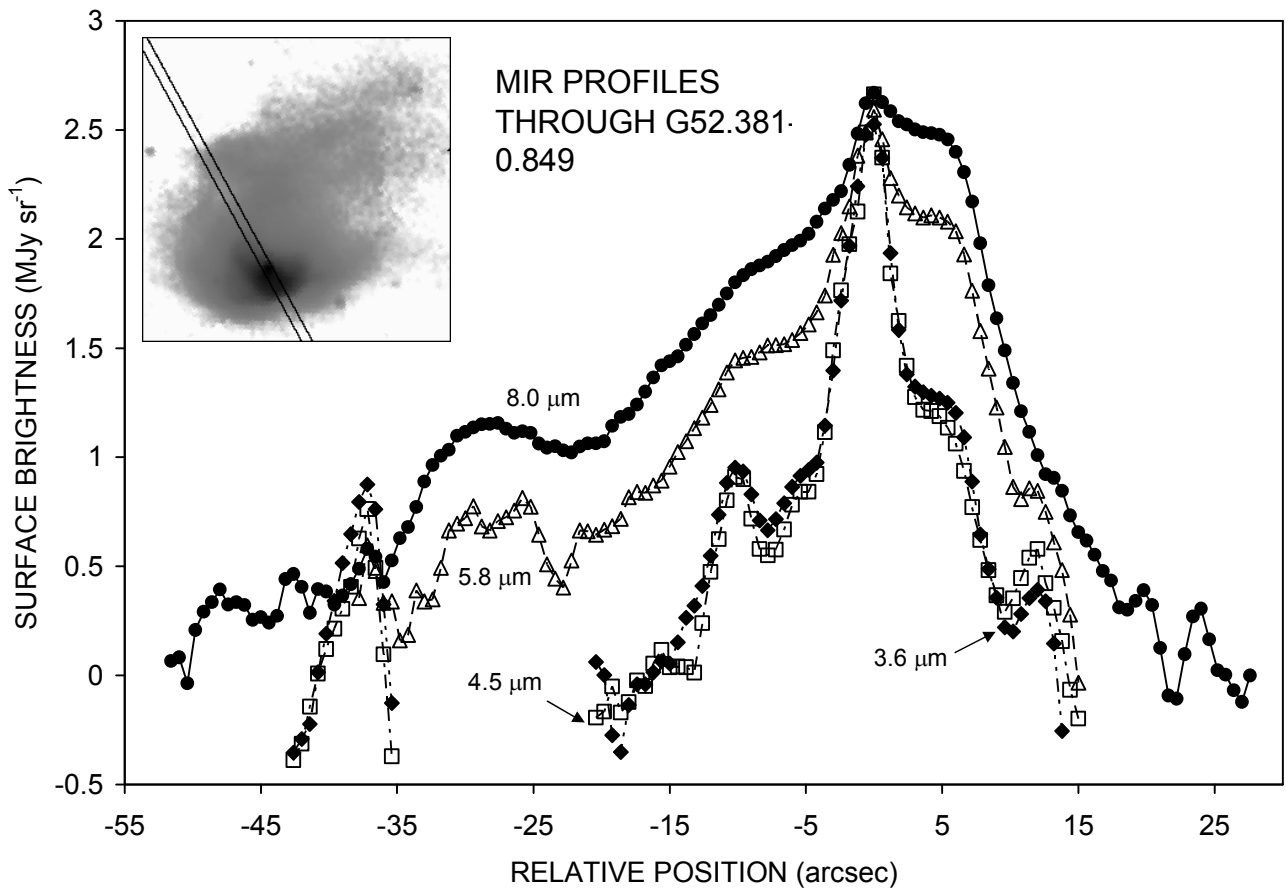

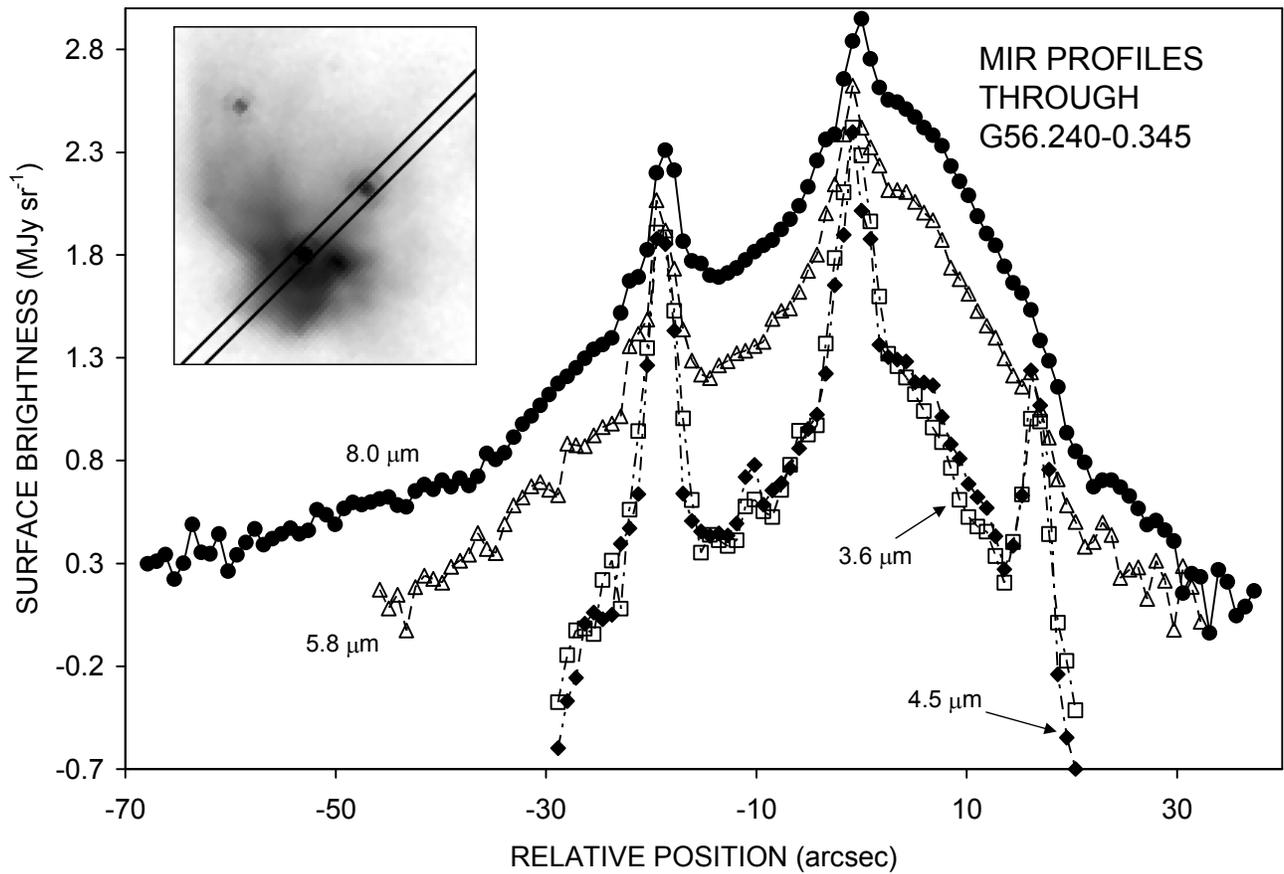

FIGURE 3

SPECTRA OF THE SYMBIOTIC
CANDIDATE STARS IPHAS-
Sy745 AND IPHAS-Sy750

IPHAS-Sy745 + 5 10$^{-16}$

IPHAS-Sy750

TiO

TiO

WAVELENGTH (Angstroms)

FLUX (ergs cm$^{-2}$ A$^{-1}$ s$^{-1}$)

FIGURE 4

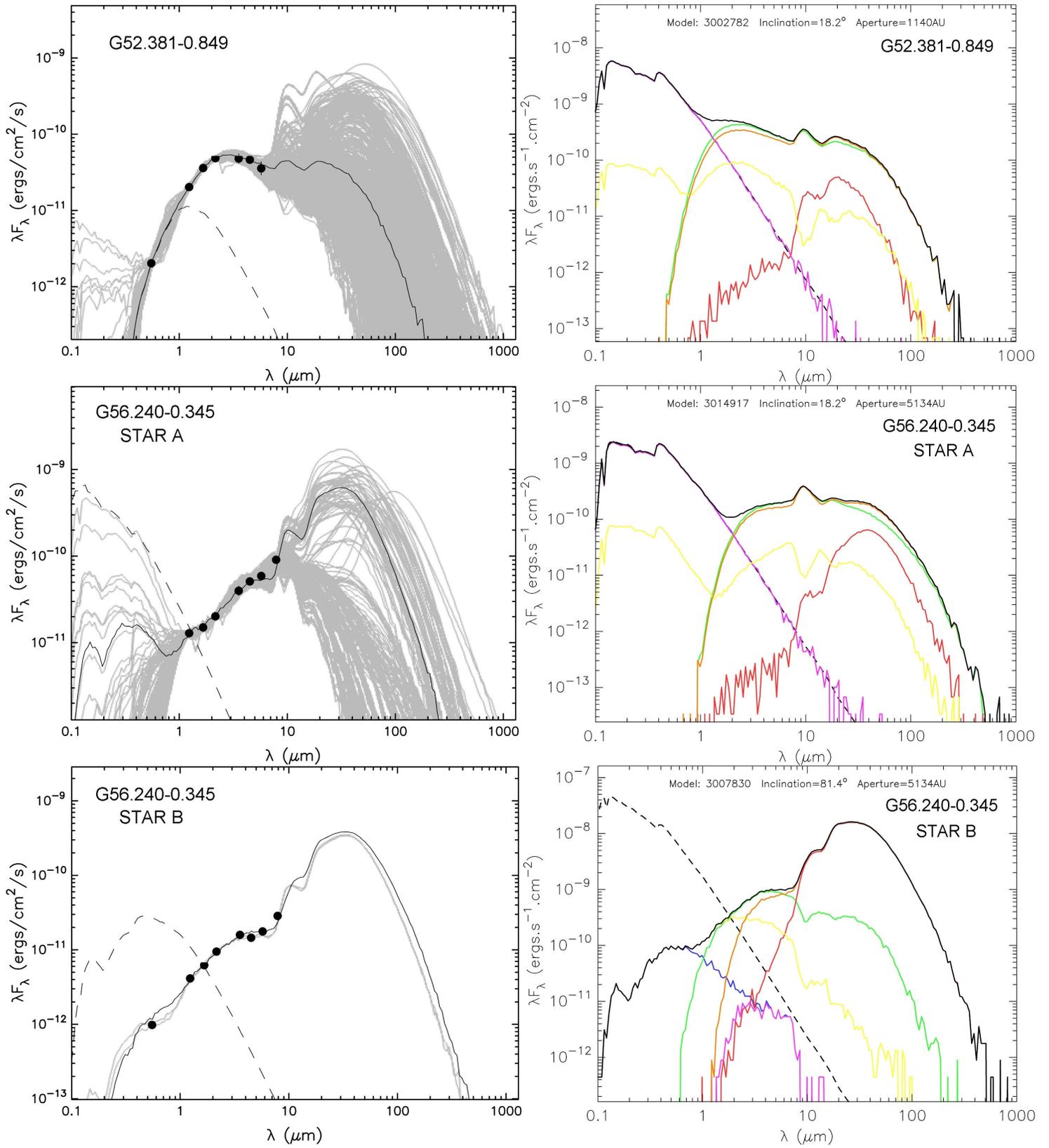

FIGURE 5

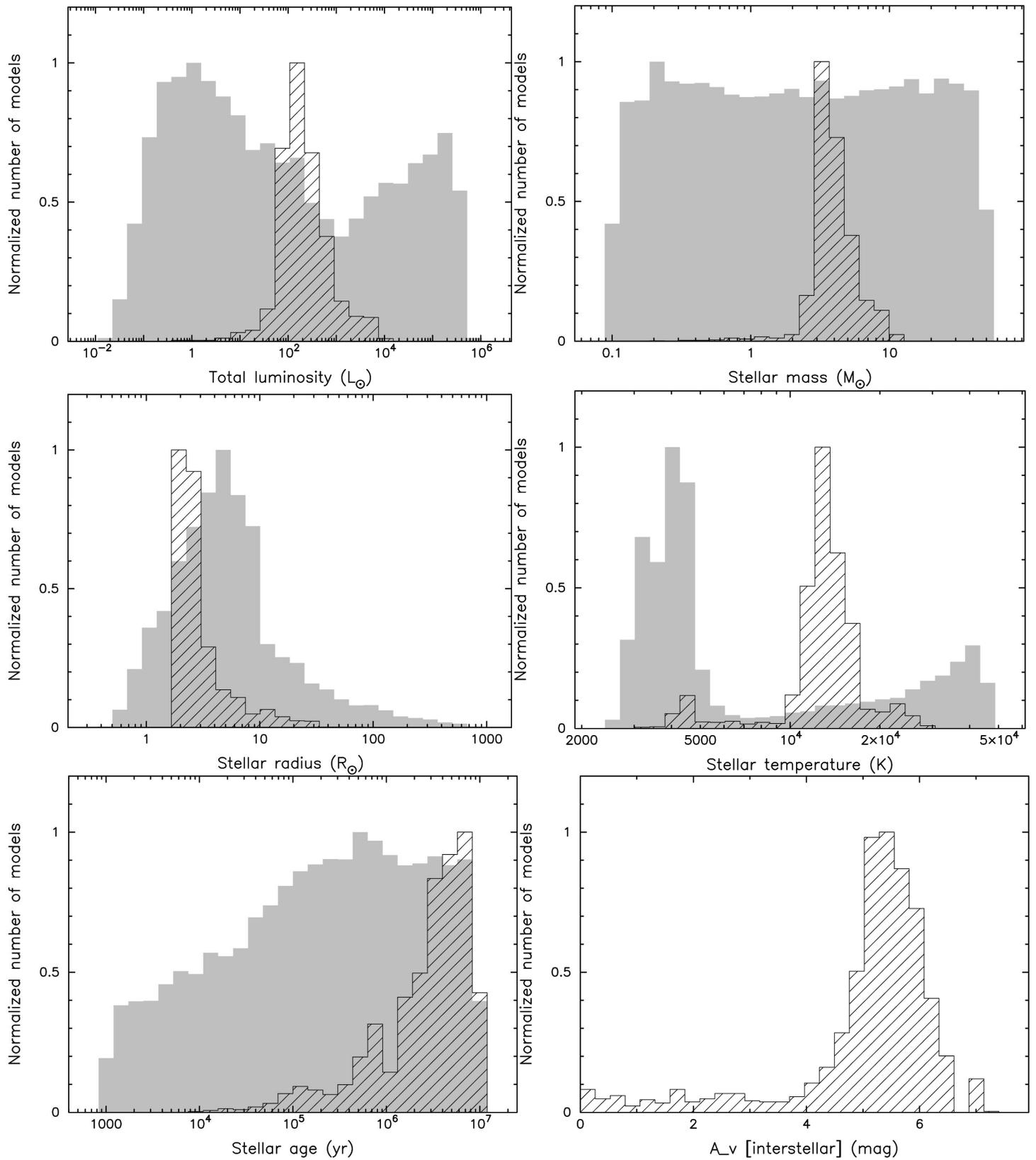

FIGURE 6

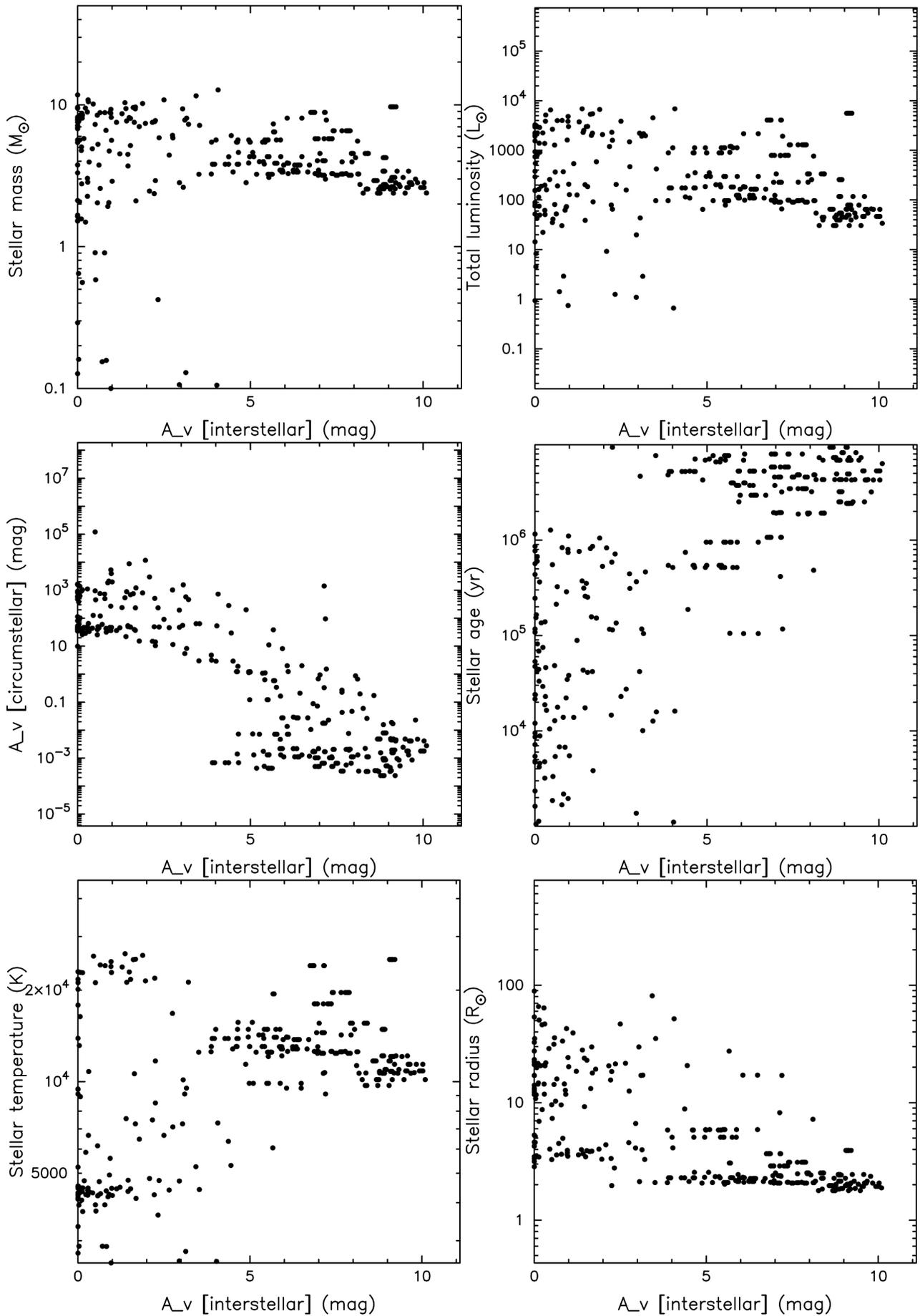

FIGURE 7